\documentstyle[preprint,eqsecnum,aps]{revtex}
\tightenlines

\newcommand{\be}{\begin{equation}}
\newcommand{\ee}{\end{equation}}
\newcommand{\bea}{\begin{eqnarray}}
\newcommand{\eea}{\end{eqnarray}}
\newcommand{\bml}{\begin{mathletters}}
\newcommand{\eml}{\end{mathletters}}
\newcommand{\pa}{\partial}

\newcommand{\bp}{\bar{\psi}}
\newcommand{\p}{\psi}
\newcommand{\T}{\Theta}
\newcommand{\dxy}{\delta(\vec{x}-\vec{y})}
\newcommand{\dxz}{\delta(\vec{x}-\vec{z})}

\newcommand{\vx}{\vec{x}}
\newcommand{\vy}{\vec{y}}

\newcommand{\vk}{\vec{k}}
\newcommand{\vu}{\vec{u}}
\newcommand{\vv}{\vec{v}}
\newcommand{\g}{\gamma}

\newcommand{\e}{\epsilon}
\newcommand{\ve}{\varepsilon}
\newcommand{\te}{\theta}
\begin{document}
\draft
\title{Canonical Quantization of the Self-Dual Model coupled to Fermions
\cite{byline}}
\author{H. O. Girotti}
\address{Instituto de F\'{\i}sica,
Universidade Federal do Rio Grande do Sul \\ Caixa Postal 15051, 91501-970  -
Porto Alegre, RS, Brazil.}

\date{March 1998}
\maketitle
\begin{abstract}
This paper is dedicated to formulate the interaction picture dynamics of
the self-dual field minimally coupled to fermions. To make this possible,
we start by quantizing the free self-dual model by means
of the Dirac bracket quantization procedure. We obtain, as result, that the
free self-dual model is a relativistically invariant quantum field theory whose
excitations are identical to the physical (gauge invariant) excitations of 
the free Maxwell-Chern-Simons theory.

The model describing the interaction of the self-dual field minimally coupled
to fermions is also quantized through the
Dirac-bracket quantization procedure. One of the self-dual field components
is found not to commute, at equal times, with the fermionic fields. Hence,
the formulation of the interaction picture dynamics is only possible after
the elimination of the just mentioned component. This procedure brings, in
turns, two new interactions terms, which are local in space 
and time while non-renormalizable by power counting. Relativistic invariance is
tested in connection with the elastic fermion-fermion scattering amplitude.     
We prove that all the non-covariant pieces in the interaction
Hamiltonian are equivalent to the covariant minimal interaction of the
self-dual field with the fermions. The high energy behavior of the self-dual
field propagator corroborates that the coupled theory is non-renormalizable.
Certainly, the self-dual field minimally coupled to fermions bears no 
resemblance with the renormalizable model defined by the Maxwell-Chern-Simons 
field minimally coupled to fermions.

\end{abstract}

\newpage
\narrowtext

\section{Introduction}
\label{sec:level1}

The self-dual (SD) model, put forward by Townsend, Pilch and Van
Nieuwenhuizen\cite{TPN1}, has been the object of several investigations.
On a semiclassical level, it has been shown to be equivalent to the
Maxwell-Chern-Simons (MCS) theory\cite{Jackiw1,Frad}. That this equivalence
holds on the level of the Green functions was proved in Ref.\cite{Rothe1}.
Lately, the second-class constraints of the SD model were successfully 
converted into first-class by means of the embedding procedure proposed by 
Batalin, Fradkin and Tyutin\cite{BFT}. It was then found that the SD model and
the MCS theory are just different gauge-fixed versions of a parent 
theory\cite{Kim,Rothe2,Rothe3}.
    
The statements above refer to the free SD and MCS models. The central purpose 
of this paper is to present the canonical quantization of the SD theory 
minimally coupled to fermions, while unraveling possible links with the MCS 
theory minimally coupled to fermions. 

Explicit calculations of probability amplitudes for an interacting field 
theory in 2+1 dimensions are, on
general grounds, possible only in perturbation theory. Our main goal
will then be to formulate the quantum dynamics of the minimally 
coupled SD-fermion system in the interaction picture. 
But to make this operational one must {\it a priori}
solve for the quantum dynamics of the free SD and fermion fields. Hence,
we start in section 2 by quantizing the uncoupled SD theory through the Dirac 
bracket quantization procedure (DBQP)\cite{Di,FVil,Su,Gi1}. The model is found
to contain only one independent 
polarization vector, which is explictly determined and turns out
to be identical to the polarization vector of the MCS in the Landau 
gauge\cite{Gi2}. By combining this result
with the field commutators at different times, obtained afterwards, we are led
to conclude that the space of states is positive definite. Ordering ambiguities
arise in the construction of the (symmetric) energy-momentum tensor and we are
forced to adopt an ordering prescription. The components of the energy-momentum
tensor are shown to fulfill the Dirac-Schwinger algebra, which secures the
existence of a set of charges obeying the Poincar\'e algebra\cite{DiSch,Sch}.
The summary of this section is that
the SD model is a relativistically invariant quantum field theory 
describing massive quanta whose spin can be $\pm 1$ depending
upon the sign of the mass parameter. Thus, the particle content of the 
SD model is identical to that of the MCS theory in the Coulomb
gauge\cite{Gi2,Ha1,Jackiw2}. This establishes the equivalence of these two 
models within the operator approach. We emphasize that the developments in this
section do not duplicate those in 
Refs.\cite{Jackiw1,Rothe1,Kim,Rothe2,Rothe3}, which
were carried out either within the semiclassical approximation\cite{Jackiw1} or
within the functional approach\cite{Rothe1,Kim,Rothe2,Rothe3}.    

The quantization of the free fermion field through the DBQP can be found in 
the literature\cite{Su,Fleck} and leads to standard results quoted in 
texbooks on quantum field theory. We then turn in section 3 into applying
the DBQP for quantizing the SD model minimally coupled to fermions. 
Aside from technicalities which will be discussed in detail, the main point
arising along this process of quantization is that the
equal-time commutators (ETC's) involving the fermions and one of the SD 
field components do not vanish,
thus obstructing the standard way for arriving at the formulation of the
quantum dynamics in the interaction picture. The situation resembles that
encountered when quantizing the MCS theory minimally coupled to fermions 
in the Coulomb gauge. There, the ETC's involving the fermionic fields and the 
longitudinal component of the momenta canonically conjugate to the gauge 
potentials do not vanish, also. The way out from the trouble consists in 
getting rid off these longitudinal components by means of the Gauss
law constraint. This operation brings, in turns, two new interaction terms to 
the Hamiltonian\cite{Gi2,fo1}. In the present case, the elimination of the 
corresponding SD field variable also brings two new interaction terms, being 
both local in space-time and non-renormalizable by power counting. One of these 
terms is the time component
of a Pauli interaction  while the other is the time component of a Thirring 
interaction. After having determined all the vertices of the coupled theory
we end this section by presenting the expressions for the free fermionic and 
SD-field propagators.

The relativistic invariance of the interacting model is tested in section 4.
As in Ref.\cite{Gi2}, we use, for this purpose, the lowest order 
contribution to the elastic fermion-fermion scattering amplitude. We 
demonstrate that the combined action of the
non-covariant pieces that make up the interaction Hamiltonian can, after all,
be replaced by the minimal covariant field-current interaction. The high 
energy behavior of the self-dual field
propagator corroborates that the SD field when minimally coupled to fermions
defines a non-renormalizable theory.  

Some final remarks and the conclusions are contained in section 5.

\section{Quantization of the free SD model}
\label{sec:level2}

The dynamics of the SD theory is described by the Lagrangian
density\cite{TPN1,Jackiw1} 

\be
\label{201}
{\cal L}^{SD}\,=\,- \frac{1}{2m} \epsilon^{\mu\nu\rho}\,
(\partial_{\mu} f_{\nu})\,
f_{\rho}\,+\,\frac{1}{2}\,f^{\mu}f_{\mu}\,\,\,,
\ee

\noindent
where $m$ is a parameter with dimensions of mass. We use natural units 
($c=\hbar=1$) and our metric is $g_{00}=-g_{11}=-g_{22}=1$. The fully 
antisymmetric tensor is normalized such that $\epsilon^{012}=1$ and we define 
$\epsilon^{ij}\equiv\epsilon^{0ij}$. Repeated Greek indices sum from $0$ to $2$
while repeated Latin indices sum from $1$ to $2$.

By computing the momenta ($\pi_{\mu}$) canonically conjugate to the field
variables ($f^{\mu}$) one finds that the theory possesses the primary
constraints 

\bml
\label{202}
\bea
&&P_0\,=\,\pi_0\,\approx\,0\,\,\,, \label{mlett:a202} \\
&&P_i\,=\,\pi_i\,+\,\frac{1}{2m} \epsilon_{ij} f^j\,\approx\,0\,\,,i=1,2\,\,\,,
\label{mlett:b202}
\eea
\eml

\noindent
where the sign of weak equality ($\approx$) is used in the sense of
Dirac\cite{Di}. Furthermore, the canonical Hamiltonian ($H^{SD}$) is found to 
read

\be
\label{203}
H^{SD}\,=\,\int d^2x\,\left(-\frac{1}{2}f^{\mu}f_{\mu}\,+\,\frac{1}{m} \e_{ij}
f^0 \partial^i f^j \right)\,\,\,.
\ee

\noindent
Hence, the total Hamiltonian ($H^{SD}_T$) is given by\cite{Di}
$H^{SD}_T = H^{SD} + \int d^3x \left( u^0 P_0 + u^i P_i \right)$, where the 
$u$'s are Lagrange multipliers. 

The persistence in time of $P_0\approx0$,

\[
\dot{P_0}\,=\,[P_0\,,\,H^{SD}_T]_P\,\approx\,0\,\,\,,
\]

\noindent
where $[,]_P$ denotes the Poisson bracket, leads to the existence of the
secondary constraint

\be
\label{204}
S\,=\,\frac{1}{m}\left(f^0\,-\,\frac{1}{m}\epsilon_{ij}\partial^if^j\right)
\,\approx\,0\,\,\,.
\ee

\noindent
On the other hand, $[P_i(\vx),P_j(\vy)]_P \neq 0$ implies that by demanding 
persistence in time of $P_i\approx 0, i=1,2$, one determines the Lagrange 
multipliers $u^i, i=1,2$. 
Similarly, $[P_0(\vx),S(\vy)]_P \neq 0$, together with the persistence in time 
of $S(\vx)\approx 0$, enables one to determine $u^0$. 

From the above analysis follows that all constraints are 
second-class and, therefore, Dirac brackets  with respect to them can be 
introduced in the usual manner\cite{Di}. The phase-space variables are,
afterwards, promoted to operators obeying an equal-time
commutator algebra which is to be abstracted from the corresponding Dirac
bracket algebra, the constraints and gauge conditions thereby translating 
into strong operator relations. This is the DBQP, which presently 
yields\cite{f2,f3}

\bml
\label{205}
\bea
&&[f^0(\vx),f^j(\vy)]\,=\,i\, \partial^j_x \dxy\,\,\,,\label{mlett:a205}\\
&&[f^k(\vx),f^j(\vy)]\,=\,-im\, \epsilon^{kj}\, \dxy\,\,\,,\label{mlett:b205}\\
&&[f^0(\vx),\pi_k(\vy)]\,=\,-\frac{i}{2m}\,\epsilon_{kj}\,
\partial^j_x \dxy\,\,\,,\label{mlett:c205}\\
&&[f^j(\vx),\pi_k(\vy)]\,=\,\frac{i}{2}\, g^j_k\, \dxy\,\,\,,
\label{mlett:d205}\\
&&[\pi_j(\vx),\pi_k(\vy)]\,=\,-\frac{i}{4m}\,\epsilon_{jk}\,
\dxy\,\,\,,\label{mlett:e205}
\eea
\eml

\noindent
whereas all other ETC's vanish. As for the quantum mechanical Hamiltonian
($H^{SD}$) it can be read off directly from (\ref{203}), in view of the 
absence of ordering ambiguities in the classical-quantum transition. This is 
true although $f^0$  and $f^j$ do not commute at equal
times (see(\ref{mlett:a205})). Alternative expressions for the operator 
$H^{SD}$ can be obtained after recalling that, within the algebra (\ref{205}),
the constraints act as strong identities. Thus,

\be
\label{206}
H^{SD}\,=\,\int d^2x\,\left(-\frac{1}{2}f^{\mu}f_{\mu}\,+\,\frac{1}{m} \epsilon_{ij}
f^0 \partial^i f^j \right)\,=\,
\int d^2x\,\left(-\frac{1}{2}f^{\mu}f_{\mu}\,+\,f^0f_0  \right)\,\,\,.
\ee

What we need next is to solve the equations of motion 

\be
\label{207}
\partial_0 f^0(\vx)\,=\,i\,[H^{SD} , f^0(\vx)]\,=\,-\partial_jf^j(\vx) 
\Longrightarrow \,\partial_{\mu}f^{\mu}\,=\,0\,\,\,,
\ee

\noindent
\be
\label{208}
\partial_0f^j(\vx)\,=\,i\,[H^{SD} , f^j(\vx)]\,=\,\partial^jf_0(\vx)\,+\,m\,
\epsilon^{jk}\,f_k(\vx)\,\,\,,
\ee

\noindent
obeyed by the field operators $f^{\mu}$. Notice that this last equation and 
(\ref{204}) can be unified into the single covariant expression 

\be
\label{209}
\epsilon^{\mu\sigma\alpha}\,\partial_{\sigma} f_{\alpha}\,-\,m\,f^{\mu}\,
=\,0\,\,\,,
\ee

\noindent
which is formally identical to the Lagrange equation of motion deriving from
(\ref{201}). Of course, (\ref{207}) follows from (\ref{209}). The solving of
(\ref{209}) is greatly facilitated if one observes that its solutions
also fulfill the Klein-Gordon equation

\be
\label{210}
\left(\Box\,+\,m^2\right)f^{\mu}\,=\,0\,\,\,,
\ee

\noindent
the converse not being necessarily true. One can then write

\be
\label{211}
f^{\mu}(x)\,=\,\int d^2z\,\Delta(x-z)\,\partial_0^z f^{\mu}(z)\,+\,
\int d^2z\,\partial_0^x \Delta(x-z)\,f^{\mu}(z)\,\,\,,
\ee

\noindent
where the 2+1 dimensional Pauli-Jordan delta function ($\Delta$) of mass $m$ 
verifies, as known, $\left(\Box_x+m^2\right)\Delta(x-z)=0$ together with the 
initial conditions $\Delta(x-z)|_{x^0=z^0}=0$ and 
$\partial_0^x \Delta(x-z)|_{x^0=z^0}=\dxz$. The velocities, in the right hand
side of (\ref{211}), can be eliminated in favor of the phase-space variables 
by using Eqs.(\ref{207}) and (\ref{208}). Once this has been done, the 
commutators at different times can be computed by using, as input, the equal 
time algebra (\ref{205}). In this way one obtains

\be
\label{212}
[f^{\mu}(x)\,,\,f^{\nu}(y)]\,=\,i\,\left(m^2\,g^{\mu\nu}\,+\,\partial^{\mu}_x
\partial^{\nu}_x\,-\,m\,\epsilon^{\mu\nu\rho}\,\partial^x_{\rho}\right)
\Delta(x-y)\,\,\,,
\ee

\noindent
which is easily seen to satisfy the first order differential equation
  
\be
\label{213}
\epsilon_{\alpha\beta\mu}\,\partial^{\beta}_x\,[f^{\mu}(x)\,,\,f^{\nu}(y)] \,
-\,m\,[f_{\alpha}(x)\,,\,f^{\nu}(y)]\,=\,0\,\,\,.
\ee

\noindent
Thus, the field configurations entering the commutator (\ref{212}) are, as
required, solutions of Eq.(\ref{209}). Since the Pauli-Jordan delta function
admits a decomposition into positive and negative frequency parts\cite{f4}, we
learn from (\ref{212}) that the same is true for the field operators 
$f^{\mu}$, namely,

\be
\label{214}
f^{\mu}(x)\,=\,f^{\mu(+)}(x)\,+\,f^{\mu(-)}(x)\,\,\,,
\ee

\noindent
with

\be
\label{215}
f^{\mu(\pm)}(x)\,=\,\frac{1}{2\pi} \int \frac{d^2k}{\sqrt{2\omega_{\vk}}} 
\exp \left[\pm i(\omega_{\vk} x^0 - \vk \cdot
\vx)\right]\,f^{\mu(\pm)}(\vk)\,\,\,, 
\ee

\noindent
$\omega_{\vk} \equiv +\sqrt{|\vk|^2+m^2}$, and

\bml
\label{216}
\bea
&&[f^{\mu (+)}(\vk),f^{\nu (+)}({\vk}\prime)]\,= \,
[f^{\mu (-)}(\vk),f^{\nu (-)}({\vk}\prime)]=0\,\,\,,\label{mlett:a216}\\ 
&&[f^{\mu(-)}(\vk)\,,\,f^{\nu(+)}({\vk}\prime)]\,=\,-\left( m^2\,g^{\mu\nu}\,
-\,k^{\mu}k^{\nu}\,+\,im\,\epsilon^{\mu\nu\rho}\,k_{\rho} \right)
\delta(\vk\,-\,{\vk}\prime)\,\,\,.\label{mlett:b216}
\eea
\eml

To go further on we must recognize that not all variables spanning
the phase are independent variables. Indeed, the system under analysis
possesses three coordinates 
($f^{\mu},\mu=0,1,2$), three momenta ($\pi_{\mu}, \mu=0,1,2$) and four 
constraints ($P_0, P_1, P_2 ,S$). Thus, we are left with only one 
independent degree of freedom which implies that there is only one 
polarization vector in the theory, to be designated by 
$\ve^{\mu}(\vk)$. We can then write

\bml
\label{217}
\bea
&&f^{\mu (+)}(\vk)\,=\,\ve^{\mu}(\vk)\,a^{(+)}(\vk)\,\,\,,\label{mlett:217a}\\
&&f^{\mu (-)}(\vk)\,=\,\ve^{\ast\mu}(\vk)\,a^{(-)}(\vk)\,\,\,,
\label{mlett:217b}
\eea
\eml

\noindent
where $a^{(\pm)}(\vk)$ are operators whose commutator algebra will be 
determined later on. By going back with (\ref{217}) into (\ref{215}) 
and with this into (\ref{209}) one finds that $\ve^{\mu}$ verifies the 
homogeneous equation

\be
\label{218}
\Sigma^{\rho\alpha}(\vk)\,\ve_{\alpha}(\vk)\,=\,0\,\,\,,
\ee

\noindent
with   

\be
\label{219}
\Sigma^{\rho\alpha}(\vk)\,\equiv\,i\,\e^{\rho\sigma\alpha}\,k_{\sigma}\,-\,
m\,g^{\rho\alpha}\,\,\,.
\ee

\noindent
The vanishing of the determinant of the matrix $\Sigma^{\rho\alpha}$ is a
necessary and sufficient condition for Eq.(\ref{218}) to have solution 
different from the trivial one. The computation of this determinant is 
straightforward and yields

\be
\label{220}
\det \| \Sigma \|\,=\,m\,(k^2\,-\,m^2)\,\,\,.
\ee

\noindent
Therefore, the theory only propagates particles with 
mass equal to $|m|$. To solve (\ref{218}) we adopt the 
strategy presented in Ref.\cite{Gi2}, i.e., we start by relating 
$\ve^{\mu}(\vk)$ with $\ve^{\mu}(0)$ through the corresponding Lorentz 
transformation, namely,

\bml
\label{221}
\bea
\ve^{0}(\vk)\,& = &\, \frac{1}{|m|}\, \vk \cdot \vec{\ve}(0), 
\label{mlett:a221} \\
\ve^{j} (\vk)\,& = &\,\ve^{j}(0)\,+\,
\frac{\vec{\ve}(0) \cdot \vk}{(\omega_{\vk}\,+\,|m|)\,|m| }
\,k^{j}\,\,\,. \label{mlett:b221}
\eea
\eml  

\noindent
This form of the solution is particularly appealing because it shows,
explicitly, that $\ve^{\mu}(\vk)$ goes continuously to the 
corresponding value in the rest frame of reference. Now, from (\ref{218}) one  
finds that

\bml
\label{222}
\bea
\ve^{0}(0)\,& = &\,0\,\,\,,\label{mlett:a222}\\
\ve^{j}(0)\,& = &\,-i\,\frac{|m|}{m}\,\e^{jl}\,\ve_{l}(0)\,\,\,,
\label{mlett:b222}
\eea
\eml
  
\noindent
which completes the determination of the polarization vector. One can check
that 

\bml
\label{223}
\bea
\ve^{\mu}(\vk) \ve_{\mu}(\vk)
\,& = &\,-{\vec{\ve}}(0) \cdot {\vec{\ve}}(0)\,=\,0, 
\label{mlett:a223} \\
\ve^{\mu}(\vk) \ve^{\ast}_{\mu}(\vk)
\,& = &\,-{\vec{\ve}}(0) \cdot {\vec{\ve}}^{\ast}(0)\,=\,
-2\,|\ve^{1}(0)|^{2}\label{mlett:b223},
\eea
\eml

\noindent
where $|\ve^{1}(0)|$ is to be fixed by normalization. This is just the
polarization vector of the MCS theory in the Landau gauge\cite{Gi2}. 
 
By substituting Eq.(\ref{217}) into Eq.(\ref{mlett:b216}) one arrives to   

\be
\label{224}
\ve^{\ast\mu}(\vk) \ve^{\nu}(\vk\prime)[a^{(-)}(\vk)\,,\,a^{(+)}(\vk\prime)]\,
=\,-\left( m^2\,g^{\mu\nu}\,
-\,k^{\mu}k^{\nu}\,+\,im\,\epsilon^{\mu\nu\rho}\,k_{\rho} \right)
\delta(\vk\,-\,{\vk}\prime)\,\,\,.
\ee

\noindent
On the other hand, if we fix $|\ve^{1}(0)|=m$ the polarization vector can be
seen to fulfill the relationship

\be
\label{225}
\ve^{\ast\mu}(\vk) \ve^{\nu}(\vk) \,=\,-\left( m^2\,g^{\mu\nu}\,
-\,k^{\mu}k^{\nu}\,+\,im\,\epsilon^{\mu\nu\rho}\,k_{\rho} \right)\,\,\,.
\ee

\noindent
By combining Eqs.(\ref{224}) and (\ref{225}) one obtains the commutator algebra
obeyed by the operators $a^{(\pm)}(\vk)$, i.e., 

\be
\label{226}
[a^{(-)}(\vk)\,,\,a^{(+)}(\vk\prime)]\,=\,\delta(\vk\,-\,{\vk}\prime)\,\,\,.
\ee

\noindent
Thus, $a^{(-)}(\vk)$ and ($a^{(+)}(\vk)$) are, respectively, destruction and 
creation operators and the space of states is a Fock space with positive
definite metric.

What remains to be done is to investigate whether the DBQP has, for the case of
the SD model, led to a relativistically invariant quantum theory and to
determine the spin of the corresponding particle excitations. The first of 
these questions appears to be
trivial because the Lagrangian density (\ref{201}) is a Lorentz scalar.
However, one is to observe that the space-time and the space-space
components of the classical symmetric (Belinfante) energy-momentum tensor 
derived in Ref.\cite{Jackiw1}   
($\Theta^{\mu\nu}\,=\,f^{\mu}f^{\nu}\,-\,\frac{1}{2}\,g^{\mu\nu}\,f^{\alpha}
f_{\alpha}$),
become afflicted by ordering ambiguities when promoted to the quantum level
(Recall the equal-time commutation algebra (\ref{205})). To cope with this
problem we introduce the symmetric ordering prescription ($A\cdot B =
1/2(AB + BA)$) and take as the Poincar\'e densities of the quantum theory the
composite Hermitean operators 

\be
\label{227}
\Theta^{\mu\nu}\,=\,f^{\mu} \cdot f^{\nu}\,
-\,\frac{1}{2}\,g^{\mu\nu}\,f^{\alpha} \cdot f_{\alpha} \,\,\,.
\ee

\noindent
After some algebra, one corroborates that the Dirac-Schwinger 
algebra\cite{DiSch,Sch},

\bml
\label{228}
\bea
& &[\T^{00}(x^{0}, \vec{x})\,,\,\T^{00}(x^{0}, \vec{y})]\,
=\,-i\,\left(\T^{0k}(x^{0}, \vec{x})\,
+\,\T^{0k}(x^{0},\vec{y})\right) \partial^{x}_{k}
\delta(\vec{x}-\vec{y}), \label{mlett:a228} \\
& &[\T^{00}(x^{0}, \vec{x})\,,\,\T^{0k}(x^{0},\vec{y})]\,=\,
-i\,\left(\T^{kj}(x^{0},\vec{x})\,
-\,g^{kj}\,\T^{00}(x^{0},\vec{y})\right)
\partial^{x}_{j}\delta(\vec{x}-\vec{y}), \label{mlett:b228} \\
& &[\T^{0k}(x^{0},\vec{x})\,,\,\T^{0j}(x^{0},\vec{y})]\,=\,-i\,
\left(\T^{0k}(x^{0},\vec{x})\partial_{j}^{x}\,+\,
\T^{0j}(x^{0},\vec{y})\partial^{x}_{k} \right) \delta(\vec{x}-\vec{y}),
\label{mlett:c228}
\eea
\eml

\noindent
holds. Then, the generators of space-time translations
($P^{\mu}$), Lorentz boosts ($J^{0i}$), and spatial rotations ($J$),

\bml
\label{229}
\bea
& &P^{0}\,\equiv\,\int d^2x\,\T^{00}(x^{0},\vx) \,=\,H^{SD}, 
\label{mlett:a229} \\
& &P^{i}\,\equiv\,\int d^2x\,\T^{0i}(x^{0},\vx), 
\label{mlett:b229} \\
& &J^{0i}\,\equiv\,- x^{0}P^{i}\,+\,\int d^2x\,
[x^{j}\T^{00}(x^{0},\vx)], \label{mlett:c229} \\
& &J\,\equiv\,\epsilon_{lj} \int d^{2}x \,x^{l} \T^{0j}(x^{0},\vx),
\label{mlett:d229} 
\eea
\eml

\noindent
fulfill the Poincar\'e algebra. To write the Poincar\'e
generators in terms of the operators $a^{(\pm)}(\vk)$ involves a 
cumbersome calculation whose details will not be given here. We 
only mention that the functional form of the polarization vector 
$\ve^{\mu}(\vk)$ (see Eqs.(\ref{221}) and (\ref{222})) plays a central role 
for arriving at the following expressions

\bml
\label{230}
\bea
& &P^{0}\,= \,H^{SD}\,=\,\int d^2k\,\omega_{\vk}\,a^{(+)}(\vk)a^{(-)}(\vk)\,\,\,,
\label{mlett:a230}\\
& &P^{j}\,= \,\int d^2k\, k^j\,\,a^{(+)}(\vk)a^{(-)}(\vk)\,\,\,,
\label{mlett:b230}\\
& &J\,=\,\frac{m}{|m|} \int d^{2}k\,
a^{(+)}(\vk)a^{(-)}(\vk)\,+\,i \e_{jl} \int d^{2}k\,
a^{(+)}(\vk)\,k^{j} \frac{\partial}{\partial k_{l}}\, a^{(-)}(\vk)\,\,\,.
\label{mlett:c230}
\eea
\eml

\noindent
By acting with $P^{\mu}$ on the one particle state $a^{(+)}(\vk)|0>$ one learns
that $a^{(+)}(\vk)$ creates particles with three-momentum $k^{\mu}$
($k^0=\omega_{\vk}$, $k^j$). The action of the operator $J$ on a single 
particle state can also be readily derived. In particular, for the rest frame 
of reference one finds that

\be
\label{231}
J\,\{ a^{+}(\vk = 0)|0> \} \,
=\,\frac{m}{|m|} \{ a^{+}(\vk = 0)|0> \},
\ee

\noindent
which sais that the spin of the SD quanta is $\pm 1$ depending upon the
sign of the mass factor.

The quantization of the free SD model is complete. We have
demonstrated that the theory only contains particles of mass $|m|$ and spin
$m/|m|$. This is just the particle content of the MCS theory 
in the Coulomb (physical) gauge\cite{f5}. On the other hand, the 
polarization vector in the SD theory is that of the MCS model
in the Landau gauge. To summarize, the SD model describes the gauge 
invariant sector of the MCS model. Therefore, these theories are quantum 
mechanically equivalent.    

\section{The SD field minimally coupled to Fermions}
\label{sec:level3}

We now bring the fermions into de game\cite{f6}. The use of Dirac's
method\cite{Di,FVil,Su,Gi1} enables one to conclude that the dynamics of
the SD model minimally coupled to fermions is described by the canonical
Hamiltonian

\bea
\label{301}
H_1\,=\,\int d^2x &&\left[\frac{i}{2}\,(\pa_k\bp)\g^k\p\,-\,     
\frac{i}{2}\,\bp \g^k (\pa_k \p)\,+\,M\,\bp\p\,-\,\frac{g}{m}\,\bp\g_k\p f^k
\right. \nonumber\\
 &-& \left. \frac{1}{2} f_{\mu}f^{\mu}\,
+\,f^0\,\left(\frac{1}{m} \e_{ij}\pa^if^j\,-
\,\frac{g}{m}\,\bp\g_0\p\right) \right]\,\,\,,
\eea

\noindent
the primary constraints

\bml
\label{302}
\bea
&&P_0\,=\,\pi_0\,\approx\,0\,\,\,, \label{mlett:a302} \\
&&P_i\,=\,\pi_i\,+\,\frac{1}{2m} \epsilon_{ij} f^j\,\approx\,0,\,\,i=1,2\,\,\,,
\label{mlett:b302}\\
&&\te_a\,=\,\pi_{\bp_a}\,-\,\frac{i}{2} \g^0_{ab}\p_b\,\approx\,0,\,\,a=1,2
\,\,\,,\label{mlett:c302}\\
&&{\bar{\te}}_a\,=\,\pi_{\p_a}\,-\,\frac{i}{2} \bp_b \g^0_{ba}\,\approx\,0,\,\,
a=1,2 \,\,\,,\label{mlett:d302}
\eea
\eml

\noindent
and the secondary constraint

\be
\label{303}
S^F\,=\,\frac{1}{m}\left(f^0\,-\,\frac{1}{m}\epsilon_{ij}\partial^if^j
\,+\,\frac{g}{m}\bp\g_0\p \right)
\,\approx\,0\,\,\,,
\ee

\noindent
where $\pi_{\p_a}$ and $\pi_{\bp_a}$ are the momenta canonically conjugate to
$\p_a$ and $\bp_a$, respectively. Furthermore, all constraints are 
second-class, as can be easily checked. 

The next step consists in introducing Dirac brackets. The direct computations
of Dirac brackets with respect to all constraints is difficult. It is easier 
to compute first partial Dirac brackets with respect to the
fermionic constraints, $\te_a $ and ${\bar{\te}}_a$, and then use this result
as input for computing the full Dirac brackets\cite{Su}. However, this
procedure is valid if and only if $\{\te_a\,{\bar{\te}}_a\} \cap \{P_0, P_j,
S^F \}=\{\phi\}$\cite{Su} which is not the case here, since $[\te_a(\vx),
S^F(\vy)]_P\neq 0$, $[{\bar{\te}}_a(\vx), S^F(\vy)]_P\neq 0$.
Nevertheless, we recall that any combination of constraints is also a 
constraint\cite{Di} and we can, therefore, replace $S^F$ by ${\tilde{S}}^F$,
which is to be constructed such that 
$[\te_a(\vx),{\tilde{S}}^F(\vy)]_P \approx 0$, 
$[{\bar{\te}}_a(\vx),{\tilde{S}}^F(\vy)]_P \approx 0$, while keeping 
$[P_0(\vx), {\tilde{S}}^F(\vy)]_P \neq 0$. It has been shown 
that\cite{Su,Fleck} that

\[
{\tilde{S}}^F(\vx)\,=\,S^F(\vx)\,+\,{\bar{\alpha}}_a(\vx)\, \te_a(\vx)\,
+\,{\bar{\te}}_a(\vx) \, \alpha_a(\vx)\approx 0\,\,\,,
\]

\noindent
with $\alpha_a = -i\frac{g}{m^2}\p_a$ and ${\bar{\alpha}}_a 
= -i\frac{g}{m^2}\bp_a$ verifies all the above requirements.

For the partial Dirac brackets with respect to the fermionic constraints
($\Delta$-brackets) one obtains

\bml
\label{304}
\bea
\left[\p_a(\vx) , \p_b(\vy)\right]_{\Delta}\,&=&\,0\,\,\,, \label{mlett:a304}\\ 
\left[\p_a(\vx) , \bp_b(\vy)\right]_{\Delta}\,&=&\,-i\,\g^0_{ab}\, \dxy \,\,\,, 
\label{mlett:b304}\\ 
\left[\bp_a(\vx) , \bp_b(\vy)\right]_{\Delta}\,&=& \,0 \,\,\,, \label{mlett:c304} 
\eea
\eml

\noindent
while all $\Delta$-brackets involving bosonic variables equal the corresponding
Poisson brackets. For any pair of functionals, $\Lambda$ and $\Omega$, of the
phase space variables, the full Dirac bracket (D-bracket) is now to be computed
as follows\cite{Su}

\be
\label{305}
[\Lambda , \Omega]_D\,=\,[\Lambda , \Omega]_{\Delta}\,-\,\sum_{q=1}^4
\sum_{p=1}^4 \int d^2u \int d^2v \,[\Lambda , \xi_p(\vu)]_{\Delta}\,R^{pq}(\vu,
\vv)\,[\xi_q(\vv) , \Omega]_{\Delta}\,\,\,,
\ee 

\noindent
where $\xi_1\equiv P_0$, $\xi_j\equiv P_j, j=1,2$, $\xi_4\equiv S^F$, 
$\parallel R\parallel \equiv \parallel Q^{-1}\parallel$ and $Q_{pq}(\vu,\vv)
\equiv [\xi_p(\vu),\xi_q(\vv)]_{\Delta}$. Notice that the fermionic 
constraints hold as strong identities within the $\Delta$-algebra
(\ref{304}) and, consequently, we can use $S^F$ instead of ${\tilde{S}}^F$ 
when computing the D-brackets through Eq.(\ref{305}).    

As in the free field case, the quantization consists in promoting all
phase-space variables to operators obeying an equal-time commutation algebra
abstracted from the corresponding D-bracket algebra. After a lengthy 
calculation one finds that the nonvanising ETC's and anticommutators read as 
follows

\bml
\label{306}
\bea
&&[f^0(\vx),f^j(\vy)]\,=\,i\, \partial^j_x \dxy\,\,\,,\label{mlett:a306}\\
&&[f^k(\vx),f^j(\vy)]\,=\,-im\, \epsilon^{kj}\, \dxy\,\,\,,\label{mlett:b306}\\
&&[f^0(\vx),\pi_k(\vy)]\,=\,-\frac{i}{2m}\,\epsilon_{kj}\,
\partial^j_x \dxy\,\,\,,\label{mlett:c306}\\
&&[f^j(\vx),\pi_k(\vy)]\,=\,\frac{i}{2}\, g^j_k\, \dxy\,\,\,,
\label{mlett:d306}\\
&&[\pi_j(\vx),\pi_k(\vy)]\,=\,-\frac{i}{4m}\,\epsilon_{jk}\,
\dxy\,\,\,,\label{mlett:e306}\\
&&[f^0(\vx),\p_a(\vy)]\,=\,\frac{g}{m}\,\p_a(\vx)\,\dxy\,\,\,,
\label{mlett:f306}\\
&&[f^0(\vx),\bp_a(\vy)]\,=\,-\frac{g}{m}\,\bp_a(\vx)\,\dxy\,\,\,,
\label{mlett:g306}\\
&&\{\p_a(\vx),\bp_b(\vy)\}\,=\,\g^0_{ab}\,\dxy\,\,\,. \label{mlett:h306}
\eea
\eml

\noindent
The Hamiltonian operator ($H$) is the quantum counterpart of $H_1$
(see Eq.(\ref{301}), i.e.,

\bea
\label{307}
H\,=\,\int d^2x &&\left[\frac{i}{2}\,(\pa_k\bp)\cdot \g^k\p\,-\,     
\frac{i}{2}\,\bp \cdot \g^k (\pa_k \p)\,+\,M\,\bp \cdot \p\,
-\,\frac{g}{m}\,\bp \cdot \g_k\p f^k
\right. \nonumber\\
 &+& \left. \frac{1}{2}\,f^0f^0\,+\, \frac{1}{2} f^{i}f^{i}\,\,\right]\,\,\,.
\eea

\noindent
Clearly, Eq.(\ref{307}) follows from Eq.(\ref{301}) after appropriate 
ordering of fermionic factors ($\p_a\cdot\bp_b \equiv 1/2 
(\p_a \bp_b - \bp_b \p_a)$). The expression for $H$ was, furthermore,
simplified by using the constraint (\ref{303}). We are entitled to do so 
because, within the algebra (\ref{306}), all constraints hold as strong 
operator relations\cite{f7}. 

The main observation, concerning the equal-time commutation algebra 
(\ref{306}),
is that the bosonic field variable $f^0$ does not commute with the fermion
fields $\p$ and $\bp$. Hence, the implementation of the quantum dynamics in the
interaction picture will only be possible after the elimination of $f^0$. This
can be done by using the constraint relation (\ref{303}) which ultimately leads
to

\be
\label{308}
H^D\,=\,H_0^D\,+\,H^D_I \,\,\,,
\ee

\noindent
where

\bea
\label{309}
H_0^D\,=\,&&\int d^2x \left[\frac{1}{2m^2}\epsilon^{ij}\,\epsilon^{kl}\,
(\partial_if^D_j)\,(\partial_kf^D_l)\,+\,\frac{1}{2}f^D_i f^D_i \right] 
\nonumber\\
+\, &&\int d^2x \left[\frac{i}{2}\,(\pa_k\bp^D)\cdot \g^k\p^D\,-\,     
\frac{i}{2}\,\bp^D \cdot \g^k (\pa_k \p^D)\,+\,M\,\bp^D \cdot \p^D \right]
\eea

\noindent
and

\bea
\label{310}
H_I^D\,=\,\int d^2x && \left[ -\,\frac{g}{m}\,\bp^D \cdot \g^k\p^D f^D_k\,-\,
\frac{g}{m^2}\,\e^{ij}\,\pa_if^D_j\,(\bp^D\cdot \g^0\p^D)\right.
\nonumber\\
&+& \left.
\frac{g^2}{2m^2}\,(\bp^D\cdot \g^0\p^D)\,(\bp^D\cdot \g^0\p^D)\right]\,\,\,.
\eea

\noindent
Here, the superscript $D$ denotes field operators belonging to the interaction
picture; the Heisenberg field operators, we were so far dealing with, bear no
picture superscript. 

Since different pictures are connected by unitary
transformations, the equal-time commutation rules obeyed by the interaction
picture field operators can be read off directly from Eq.(\ref{306}). Then,
the equation of motion obeyed by the operator $\p^D$ ($\pa_0 \p^D 
= i [H_0^D,\p^D]$) is just the free Dirac equation and the corresponding 
momentum space fermion propagator ($S(p)$) is well known to be  

\be
\label{311}
S(p)\,=\,i\,\frac{M\,+\,\g \cdot p}{p^2 - M^2 + i\e}\,\,\,.
\ee

\noindent
The equations of motion satisfied by the operators $f_i^D, i=1,2$ are exactly
those studied in detail in section 2 of this paper. Hence, from 
Eqs.(\ref{215}), (\ref{217}), (\ref{221}), (\ref{222}) and (\ref{226}) one 
finds that the momentum space Feynman propagator ($D_{lj}(k)$) is given by

\be
\label{312}
D_{lj}(k)\,=\,\frac{i}{k^2 - m^2 +i\e} \left( - m^2 g_{lj}\,+\,k_l k_j\,
-\,i m \e_{lj} k_0\right)\,=\, D_{jl}(-k)\,\,\,.
\ee
   
\noindent
To complete the derivation
of the Feynman rules, we turn into analyzing the vertices of the SD model
minimally coupled to fermions. We observe that $H_I^D$ in Eq.(\ref{310}) 
is made up by three different kind of monomial terms. The first monomial is 
the spatial part of the standard field-current interaction, whereas the second
and third 
monomials arised as by products in the process of elimination of $f_0$. 
Unlike the cases of MCS minimally coupled to fermions\cite{Gi2} and Quantum 
Electrodynamics, these extra terms are strictly local in space-time. Also,
they are non-renormalizable by power counting. The second term
in Eq.(\ref{310}) is the time component of a  magnetic coupling, while
the third one is the time component of a four-fermions (Thirring) interaction. 

The Feynman rules derived in this section are non-manifestly covariant. Then,
we must elucidate whether or not this set of rules
leads to a relativistically invariant $S$-matrix. As in section 2, 
one may first build the symmetric energy-momentum tensor for the interacting 
theory and then use the equal-time commutation rules (\ref{306}) to 
check the fulfillment of the Dirac-Schwinger algebra. However, the validity 
of such procedure would now be dubious, since the components of the 
energy-momentum 
tensor necessarily involve (ill defined) products of Heisenberg field 
operators evaluated at the same space-time point. The next section is 
dedicated to test the relativistic invariance of the coupled theory in 
connection with the specific process of elastic fermion-fermion scattering.
Since we are dealing with a non-renormalizable theory, our computations will be
restricted to the tree approximation.   

\section{Lowest order elastic fermion-fermion scattering amplitude}
\label{sec:level4}

From the inspection of Eq.(\ref{310}) follows that the contributions of order
$g^2$ to the lowest order elastic fermion-fermion scattering amplitude 
($R^{(2)}$) can be grouped into four different kind of terms,

\be
\label{401}
R^{(2)}\,=\,\sum_{\alpha=1}^{4}R^{(2)}_{\alpha},
\end{equation}

\noindent
where

\bml
\label{402}
\bea
R^{(2)}_1 & = & -\frac {g^2}{2}(\g^k)_{ab}(\g^l)_{cd}
\int d^3x \int d^3y
<\Phi _f|T \left\{ \frac{1}{m} :\bp^D_a(x)\p^D_b(x) f^D_k(x): \right.
\nonumber \\ & \times & \left.  
\frac{1}{m} :\bp^D_c(y) \p^D_d (y)f^D_l(y): \right\}|\Phi _i>
\,\,\,, \label{mlett:a402} \\
R^{(2)}_2 & = & \frac {-ig^2}{2}(\g^0)_{ab}(\g^0)_{cd}
\int d^3x\int d^3y \delta (x - y) \nonumber \\
& \times & <\Phi _f|\frac{1}{m^2}:\bp^D_a (x) \p^D_b (x)
\bp^D_c (y) \p^D_d (y):|\Phi _i>
\,\,\,, \label{mlett:b402}\\
R^{(2)}_3 & = & - g^2 (\g^k)_{ab}(\g^0)_{cd}
\int d^3x \int d^3y
<\Phi _f| T\left\{ \frac{1}{m}:\bp^D_a(x) \p^D_b(x) f^D_k(x): \right.
\nonumber  \\
& \times & \left.  :\frac{1}{m^2} \e^{li} \left(\pa_l^y f^D_i(y)\right)
\bp^D_c(y) \p^D_d(y):\right\} | \Phi _i> \,\,\,, 
\label{mlett:c402} \\
R^{(2)}_4 & = & -\frac {g^2}{2} (\gamma ^0)_{ab} (\gamma ^0)_{cd} \nonumber \\
& \times & \int d^3x \int d^3y
< \Phi _f| T\left\{ :\frac{1}{m^2} \e^{kj} \left(\pa_k^xf^D_j(x)\right)
\bp^D_a(x) \p^D_b(x): \right.
\nonumber \\ & \times & \left. :\frac{1}{m^2} \e^{li} 
\left(\pa_l^yf^D_i(y)\right)
\bp^D_c(y) \p^D_d(y):\right\} |\Phi _i>\,\,\,. \label{mlett:d402}
\eea
\eml

\noindent
Here, $T$ is the chronological ordering operator, whereas $|\Phi_i>$ and 
$|\Phi_f>$ denote the initial and final state of the reaction, respectively. 
For the case under analysis, both $|\Phi_i>$ and $|\Phi_f>$ are two-electron 
states. Fermion states obeying the free Dirac equation in
2+1-dimensions were explicitly constructed in Ref.\cite{Gi3}, where the
notation $v^{(-)}(\vec{p})\, (\bar{v}^{(+)}(\vec{p}))$ was employed to
designate the two-component spinor describing a free electron of two-momentum
$\vec{p}$, energy $p^{0}=+({\vec{p}}^{\,2}+m^{2})^{1/2}$ and spin $s= M/|M|$ in
the initial (final) state. The plane wave expansion of the free fermionic
operators $\psi$ and $\bar{\psi}$ in terms of these spinors and of the
corresponding creation and annihilation operators goes as usual. 

In terms of the initial ($p_{1}, p_{2}$) and final momenta
($p_{1}^{\prime}, p_{2}^{\prime}$), the partial amplitudes in (\ref{402}) are
found to read

\bml
\label{403}
\bea
R^{(2)}_{1} & = & \frac {1}{2\pi}\delta^{(3)}(p_1'+p_2'-p_1-p_2)
\nonumber \\
& \times & \left\{ [\bar v^{(+)} (\vec{p}_1^{\,\prime})
(ig\gamma ^j)v^{(-)}(\vec p_1)][\bar v^{(+)}(\vec{p}_2^{\,\prime})
(ig\gamma^l)v^{(-)}(\vec p_2)] \frac{1}{m^2} D_{jl}(k)\,- \, p_1^{\,\prime}
\longleftrightarrow p_2^{\,\prime} \right\}\,\,\,, \label{mlett:a403} \\
R^{(2)}_{2} & = & \frac{1}{2\pi}\delta ^{(3)}(p_1'+p_2'-p_1-p_2)
\nonumber \\
& \times & \left\{ \frac{i}{m^2}
[\bar v^{(+)}(\vec p_1^{\,\prime}) (ig\gamma ^0) v^{(-)}(\vec p_1)]
[ \bar v^{(+)}(\vec p_2^{\,\prime})(ig\gamma ^0)v^{(-)}(\vec p_2)]
\,-\, p_1^{\,\prime}
\longleftrightarrow p_2^{\,\prime} \right\}\,\,\,, \label{mlett:b403} \\
R^{(2)}_{3} & = & \frac{1}{2\pi}\delta^{(3)}(p_1'+p_2'-p_1-p_2)
\nonumber \\
& \times & \left\{ [\bar v^{(+)}(\vec p_1^{\,\prime}) (ig\gamma ^j)
v^{(-)}(\vec p_1)]
[ \bar v^{(+)}(\vec p_2^{\,\prime})(ig\gamma ^0)v^{(-)}(\vec p_2)]
\frac{1}{m^2} \Gamma_j(k) \right. 
\nonumber \\
& + & \left. [\bar v^{(+)}(\vec p_1^{\,\prime}) (ig\gamma ^0)
v^{(-)}(\vec p_1)]
[ \bar v^{(+)}(\vec p_2^{\,\prime})(ig\gamma ^j)v^{(-)}(\vec p_2)]
\frac{1}{m^2} \Gamma_j(-k)\,
- \,p_1^{\,\prime} \longleftrightarrow p_2^ {\,\prime} \right\}\,\,\,,
\label{mlett:c403} \\
R^{(2)}_{4} & = & \frac{1}{2\pi}\delta ^{(3)}(p_1'+p_2'-p_1-p_2)
\nonumber \\
& \times & \left\{ [\bar v^{(+)} (\vec p_1^{\,\prime})(ig\gamma ^0)
v^{(-)}(\vec p_1)]
[ \bar v^{(+)}(\vec p_2^{\,\prime})(ig\gamma ^0)v^{(-)}(\vec p_2)]
\frac{1}{m^2} \Lambda(k)\,
- \,\,p_1^{\,\prime} \longleftrightarrow p_2^ {\,\prime} \right\}\,\,\,,
\label{mlett:d403}
\eea
\eml

\noindent
where 

\bml
\label{404}
\bea
&& \frac{1}{m^2} D_{jl}(k)\,
=\,\frac {i}{k^2- m^2 + i\epsilon}\left(- g_{jl}\, + \, \frac{k_j k_l}
{m^2}\,-\,\frac{i}{m} \e_{jl} k_0 \right)\,\,\,, \label{mlett:a404} \\
&& \frac{1}{m^2}\Gamma_j(k)\,=\,\frac {i}{k^2- m^2 + i\epsilon }
\left(\frac{ i \e_{jl} k^l}{m} \,+\,\frac{k_j k_0}{m^2} \right)\,\,\,, 
\label{mlett:b404} \\
&&\frac{1}{m^2} \Lambda (k)\,=\,\frac{i}{k^2-\theta ^2 + i\epsilon }
\left( \frac{-k^l k_l}{m^2} \right)\,\,\,,\label{mlett:c404}
\eea
\eml

\noindent
and
 
\be
\label{405}
k \equiv  p_1'-p_1\,=\,-(p_2'-p_2)
\ee

\noindent
is the momentum transfer (our convention for the Fourier 
integral representation is
$f(x)= 1/(2\pi)^{3}\int d^{3}k f(k) \exp(- ik\cdot x)$). By substituting
(\ref{403}) into (\ref{401}) and after taking into account (\ref{404}) one
arrives to

\bea
\label{406}
R^{(2)}\,& = &\,\left( -\frac{g^2}{2\pi}\right) \delta^{(3)}(p_1'+p_2'-p_1-p_2)
\nonumber \\
& \times & \left\{ [\bar v^{(+)}(\vec p_1^{\,\prime})
\gamma^{\mu} v^{(-)}(\vec p_1)]\,[\bar v^{(+)}(\vec p_2^{\,\prime})
\gamma^{\nu} v^{(-)}(\vec p_2)]\,\frac{1}{m^2}\,D_{\mu\nu}(k)\, 
- p_1^{\,\prime} \longleftrightarrow p_2^ {\,\prime} \right\}\,\,\,,
\eea

\noindent
where

\be
\label{407}
\frac{1}{m^2}\,D_{\mu\nu}(k)\,
=\,-\,\frac{i}{k^2 - m^2 + i\e}\,\left(g_{\mu\nu}\,
+\,i\,\e_{\mu\nu\sigma} \frac{k^{\sigma}}{m} \right)\,\,\,.
\ee

The amplitude in Eq.(\ref{406}) is a Lorentz scalar. The theory has then 
passed the test on relativistic invariance. Also, one is to observe 
that $D_{\mu\nu}(k)/m^2 $ in Eq.(\ref{406}) is contracted into conserved 
currents and, hence, terms proportional to $k^{\mu}$ can be added at will. 
Thus, we can replace $D_{\mu\nu}(k)/m^2$, given at Eq.(\ref{407}), by

\be
\label{408}
\frac{1}{m^2}\,D_{\mu\nu}^F (k)\,
=\,-\,\frac{i}{k^2 - m^2 + i\e}\,\left(g_{\mu\nu}\,
-\,\frac{k_{\mu}k_{\nu}}{m^2}\,
+\,i\,\e_{\mu\nu\sigma} \frac{k^{\sigma}}{m} \right)\,\,\,,
\ee   

\noindent
which, up to contact terms, is the free field propagator arising from
(\ref{212}). In the tree approximation one is, therefore,
allowed to replace all the non-covariant terms in $H^D_I$ (see Eq.(\ref{310})
by the minimal covariant interaction 
$-\frac{g}{m} \bp^D \cdot \g^{\mu}\p^D f^D_{\mu}$. 

We close this section by noticing that the high energy behavior of
the propagator in (\ref{408}) is radically different from that of the MCS 
theory in the Landau gauge\cite{Gi2,Jackiw2},

\be
\label{409}
D_{\mu\nu}^L (k)\,
=\,-\,\frac{i}{k^2 - m^2 + i\e}\,\left(g_{\mu\nu}\,-\,\frac{k_{\mu} k_{\nu}}
{k^2} \,+\,i\,m\,\e_{\mu\nu\sigma} \frac{k^{\sigma}}{k^2} \right)\,\,\,.
\ee

\noindent
In fact, the SD model coupled to fermions is a non-renormalizable theory. 
This corroborates the arguments in the previous section, based on power
counting.

\section{Final remarks}
\label{sec:level5}

The equivalence between the free SD and MCS models was first observed, 
at the level of
the field equations, in Ref.\cite{Jackiw1}. As known\cite{Jackiw2}, 
the Lagrangian density describing the dynamics of the free MCS theory,

\be
\label{501}
{\cal L}^{MCS} = -\frac{1}{4}F_{\mu\nu}F^{\mu\nu}+\frac{m}{4}
\epsilon^{\mu\nu\alpha}F_{\mu\nu}A_{\alpha},
\ee

\noindent
where $F_{\mu\nu}\equiv \pa_{\mu}A_{\nu}-\pa_{\nu}A_{\mu}$, leads to the 
equation of motion

\be
\label{502}
\pa_{\alpha}F^{\alpha\beta}\,+\,\frac{m}{2}\,\e^{\alpha\mu\alpha}
F_{\mu\alpha}\,=\,0\,\,\,.
\ee

\noindent
It is now easy to see that the mapping 

\be
\label{503}
F^{\beta}\,\equiv\,\frac{1}{2}\,\e^{\beta\mu\alpha}
F_{\mu\alpha}\,\longrightarrow \,f^{\beta}\,\,\,,
\ee

\noindent
carries Eq.(\ref{502}) into Eq.(\ref{209}). In section 2 we proved that this
equivalence holds rigorously at the quantum level. Since the SD
field $f^{\beta}$ is identified with the vector $F^{\beta}$, dual of the tensor
$F^{\mu\nu}$, only the gauge invariant excitations of the MCS theory must 
show up in the SD model. As demonstrated in this work, this is the case. 

The proof of equivalence of the quantized versions of the free SD and MCS 
models, given in this paper, was entirely carried out within the operator 
approach. 
One the main advantages of our presentation is that relativistic invariance 
is kept operational in all stages. This is to be compared with 
the outcomes in Refs.\cite{Rothe1,Kim,Rothe2,Rothe3}. Frequently, the 
embedding of a second-class system in a larger phase-space, with the aim of
making it first-class\cite{BFT}, implies in loosing relativistic 
invariance\cite{f8}. 

The formulation of the interaction picture dynamics for the SD model minimally
coupled to fermions was possible only after the elimination
of the degree of freedom $f^0$, whose ETC's with the fermionic fields do not
vanish. As consequence, the Hamiltonian formulation did not retain any relic 
of relativistic covariance. Nonetheless, we proved
that the non-covariant pieces in $H^D_I$ are equivalent to the 
minimal covariant field-current interaction. The high energy behavior of the 
$f$-propagator signalizes that the coupled theory is non-renormalizable.   

Thus, the SD model minimally coupled to fermions bears no resemblance
with the renormalizable model defined by the MCS field minimally coupled 
to fermions. To give further support to this conclusion we recall that it 
has recently been shown\cite{Malacarne} that the SD 
model minimally coupled to fermions is equivalent to the MCS model with 
{\it non-minimal} magnetic coupling to fermions. Also, for the fermionic 
sectors of the two theories to agree one is to add a Thirring like interaction
in one of the models. These are, of course, non-renormalizable
field theories.


\begin{references}
\bibitem {fo1} A similar situation arises in the Coulomb gauge formulation of
Quantum Electrodynamics. However, in this case, the elimination of the 
longitudinal component of
the momenta brings only one new interaction term known as the instantaneous
Coulomb interaction.
\bibitem {f2} The time label in the arguments of the fields will be
omitted whenever does not lead to confusion.
\bibitem {f3} We shall not distinguish between a quantum field operator and 
its classical counterpart.
\bibitem {f5} Recall that the MCS theory in the Landau gauge exhibits massive 
and massless (gauge) excitations\cite{Gi2}.
\bibitem {f6} For the $\gamma$-matrices we adopt the representation
$\gamma^{0}=\sigma^{3},\gamma^{1}=i\sigma^{1},\gamma^{2}=i\sigma^{2}$, where
$\sigma^{i}, i=1,2,3,$ are the Pauli spin matrices. We shall denote by $M$ the
fermion mass parameter and by $g$ the coupling constant.
\bibitem {f7} The fermion factor in (\ref{303}) must also be
understood, after quantization, as appropriately ordered.
\bibitem {f8} Relativistic invariance may be regained after an appropriate
gauge election.
\end{references}
\end{document}